\documentclass[runningheads]{llncs}
\usepackage{graphicx}
\usepackage{hyperref}
\usepackage{pdfpages}

\begin{document}
\title{Information Diffusion Power of Political Party Twitter Accounts During Japan's 2017 Election}
\titlerunning{Information Diffusion Power of Political Parties}
%
\author{
Mitsuo Yoshida\inst{1}\orcidID{0000-0002-0735-1116}
\and
Fujio Toriumi\inst{2}\orcidID{0000-0003-3866-4956}}
\authorrunning{M. Yoshida and F. Toriumi}
%
\institute{
Toyohashi University of Technology, Toyohashi, Aichi, Japan \\
\email{yoshida@cs.tut.ac.jp}
\and
The University of Tokyo, Bunkyo-ku, Tokyo, Japan \\
\email{tori@sys.t.u-tokyo.ac.jp}
}
\maketitle              
\begin{abstract}
In modern election campaigns, political parties utilize social media to advertise their policies and candidates and to communicate to electorates.
In Japan's latest general election in 2017, the 48th general election for the Lower House, social media, especially Twitter, was actively used.
In this paper, we perform a detailed analysis of social graphs and users who retweeted tweets of political parties during the election.
Our aim is to obtain accurate information regarding the diffusion power for each party rather than just the number of followers.
The results indicate that a user following a user who follows a political party account tended to also follow the account.
This means that it does not increase diversity because users who follow each other tend to share similar values.
We also find that followers of a specific party frequently retweeted the tweets.
However, since users following the user who follow a political party account are not diverse, political parties delivered the information only to a few political detachment users.

\keywords{Twitter \and Information diffusion \and Political party \and Election}
\end{abstract}

\setcounter{footnote}{0}

\section{Introduction}

A partial amendment of Japan's Public Officers Election Act in 2013 authorized election campaigning using the Internet.
In Japan's latest general election in 2017, the 48th general election for the Lower House, social media, especially Twitter\footnote{\url{https://twitter.com/}}, was actively used.
The main feature of this election is that a new party named `The Constitutional Democratic Party of Japan' attracted many followers on social media\footnote{Bloomberg (2017): A 3-Day-Old Japanese Political Party Has Already Overtaken Abe's on Twitter.}.
Political parties can deliver information to their followers, but can the parties deliver to users who are not followers, political detachment users?
If the followers retweet the tweets of political parties, the parties can deliver the information to political detachment users.
Specifically, information diffusion power cannot be determined by just the number of followers.
We are interested in how many users have political parties' information delivered to them.

Tumasjan et al. reported that Twitter is functioning as a discussion forum for politics~\cite{Tumasjan2010},
and, there have been many studies on users' access to political information on social media.
Previous studies show that users are divided regarding access to political information~\cite{Adamic2005,Hayat2017,Hyun2014,Iyengar2009}.
Such studies mainly cover political information written in the news on social media.
Previous studies also show a relationship exists between the division of information and ideology~\cite{Barbera2015,Batorski2018,Dahlgren2005,Williams2015}.
In other words, the studies were analyzing users who tended to support specific political parties.
These studies focus on users who received political information.

The number of followers is used as the attention degree of the user.
The influence of the number of followers in information diffusion is not necessarily large~\cite{Cha2010}, and fraudulent methods are sometimes used to gain followers~\cite{Stringhini2013}.
Even in political communication, the hub of information cannot be determined only by the number of followers~\cite{Bakshy2011}.

In this paper, we perform a detailed analysis of social graphs and users who retweeted tweets of political parties during the 48th general election for Japan's Lower House in 2017.
Our aim is to obtain accurate information regarding the diffusion power for each party rather than just the number of followers.
To this end, we address the following three research questions:
\begin{description}
 \item[RQ1] Homogeneity of followers:\\ Who is following the user who is following a political party account?
 \item[RQ2] Activity level of followers:\\ How many tweets were retweeted by followers of a political party account?
 \item[RQ3] Information diffusion power:\\ How many tweets were delivered to political detachment users?
\end{description}
The results indicate that a user following a user who follows a political party account tended to also follow the account.
This means that it does not increase diversity because users who follow each other tend to share similar values.
We also find that followers of a specific party frequently retweeted the tweets.
However, since users following the user who follows a political party account are not diverse, political parties delivered the information only to a few political detachment users.

\section{Dataset}

\subsection{Development}

In this study, we use Japanese retweets on Twitter collected from 28 September\footnote{The Lower House in Japan was dissolved on this day.} to 23 October 2017.
This period encompasses the 48th general election for the Lower House in Japan.
The data were collected using the Twitter Search API\footnote{We constantly searched by query ``RT lang:ja''.} and consist of 42,651,648 retweets.

To focus only on data related to politics, we targeted the official accounts of political parties and selected the accounts of the six major parties.
These political parties are shown in Tables~\ref{tb:parties} and \ref{tb:accounts}.
In the collected retweets, we only use 732,861 retweets (84,043 users) in which the tweets of these political parties have been retweeted.
Normally, tweets collected using the Twitter Streaming API and the ``follow'' parameter are used in this type of study.
Since we started this study after the end of this election, we decided to extract necessary data from the collected retweet data.

We also used a social graph on Twitter.
First, we gathered users who were following the major parties.
Then, we created a set of users that combines these users and the users who have retweeted tweets of the major parties.
This set includes 460,683 users.
Finally, we gathered users who were following the 460,683 users on 10 November 2017 and build the social graph.
As a result, the social graph consists of 16,742,073 nodes (users) and 409,741,963 edges (followee-follower paths).

\begin{table}[t]
\caption{The major political parties in Japan: ``The Liberal Democratic Party of Japan'' and ``Komeito'' are the ruling party.}
\centering
\begin{tabular}{l|l}
\hline
Party Name & Screen Name \\
\hline
The Liberal Democratic Party of Japan		& @jimin\_koho \\
The Constitutional,Democratic Party of Japan	& @CDP2017 	\\
The Party of Hope						& @kibounotou 	\\
Komeito							& @komei\_koho 	\\
The Japanese Communist Party			& @jcp\_cc 	\\
The Japan Innovation Party				& @osaka\_ishin \\
\hline
\end{tabular}
\label{tb:parties}
\end{table}

\begin{table}[t]
\caption{The accounts of major political parties in Japan: The number of seats is the result of this election. There is no correlation between the number of seats and the number of followers.}
\centering
\begin{tabular}{l|rrr|r}
\hline
Screen Name & \# of tweets & \# of retweeted & \# of followers & \# of seats \\
\hline
@jimin\_koho	& 280	& 110,685	& 134,595	& 284	\\
@CDP2017	& 904	& 506,432	& 191,011	& 55	\\
@kibounotou	& 410	& 21,559	& 13,529	& 50	\\
@komei\_koho	& 289	& 31,072	& 76,743	& 29	\\
@jcp\_cc		& 347	& 48,203	& 42,508	& 12	\\
@osaka\_ishin	& 281	& 13,163	& 15,999	& 11	\\
\hline
\end{tabular}
\label{tb:accounts}
\end{table}

\subsection{Basic Statistics}

\begin{figure}[tp]
  \centering
  \includegraphics[width=0.9\linewidth]{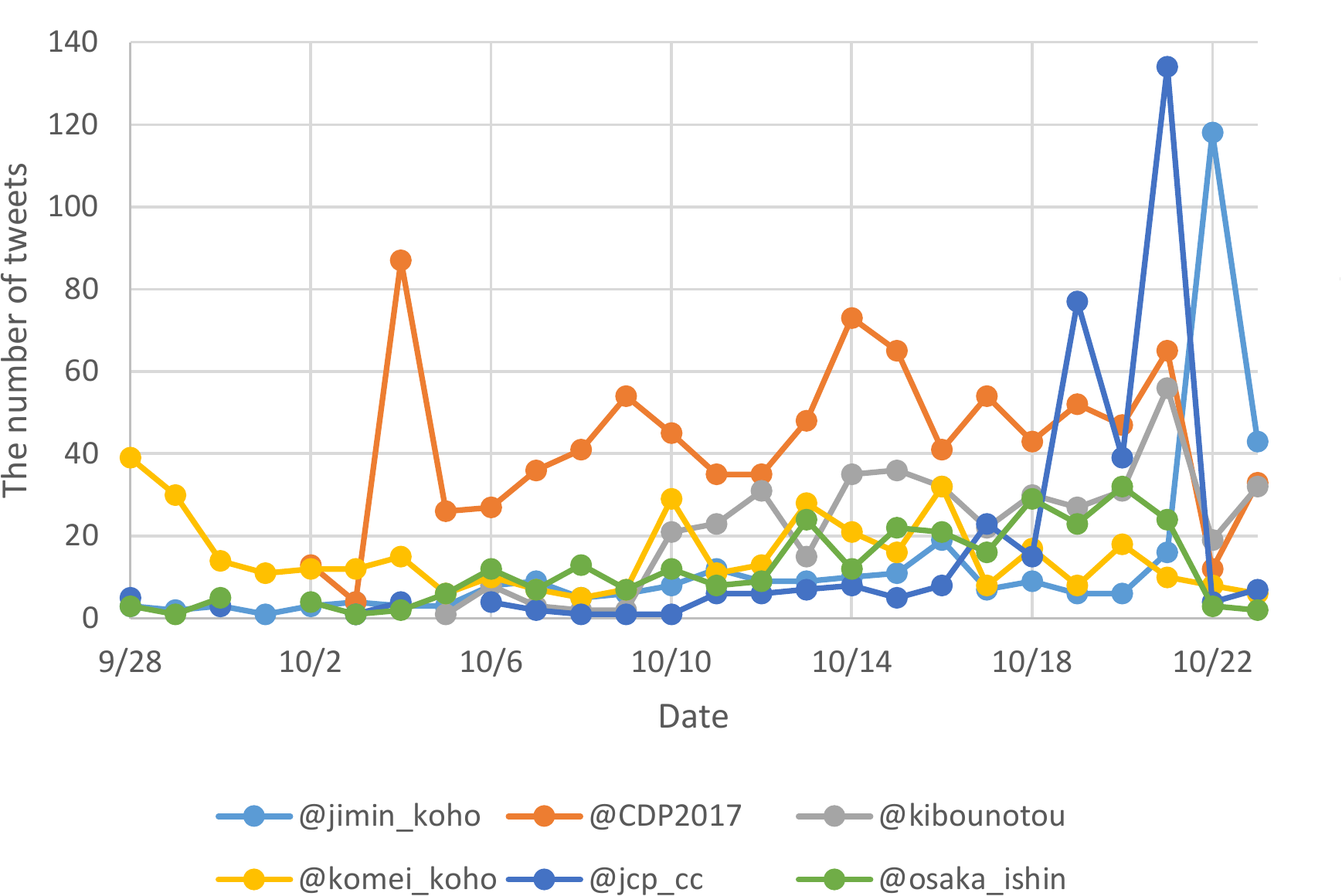}
  \caption{The number of daily tweets: the y-axis indicates the number of tweets posted by each political party. @CDP2017 tweeted daily. The number of tweets by @jcp\_cc increased on the eve of the voting day, and the number of tweets by @jimin\_koho increased on the voting day.}
  \label{fig:BS1}
\end{figure}

\begin{figure}[tp]
  \centering
  \includegraphics[width=0.9\linewidth]{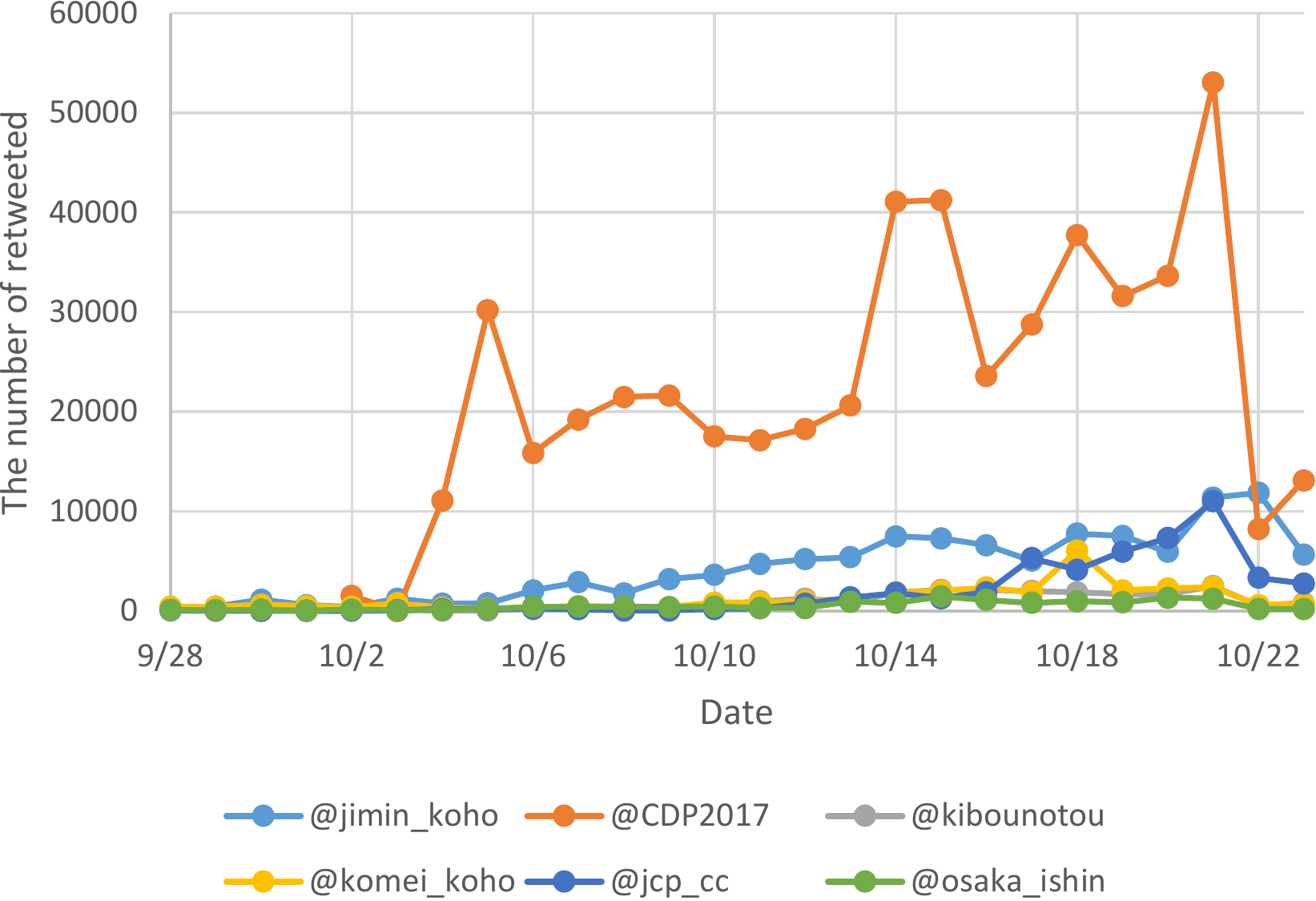}
  \caption{The number times political parties' tweets were retweeted: the y-axis indicates the retweeting frequency. Tweets of @CDP2017 were frequently retweeted.}
  \label{fig:BS2}
\end{figure}

Figures~\ref{fig:BS1} and \ref{fig:BS2} show the number of tweets for six political parties and the number times political parties' tweets were retweeted.
Under Japanese law, political parties cannot call for voting during the actual voting time (7 am to 10 pm on 22 October 2017),
which is why the number of tweets on the voting day was relatively low.
The number of tweets by @jcp\_cc increased on the eve of the voting day, and the number of tweets by @jimin\_koho increased on the voting day.
@jcp\_cc called for voting the day before the voting day, and @jimin\_koho reported each time a winner was decided.
It seems that the number of times the tweets were retweeted was greatly affected by the number of followers.
As shown in Table~\ref{tb:parties}, the number of followers of @CDP2017 is large.
As a result, the number of times tweets of @CDP2017 was retweeted increased.

Regarding information diffusion power, we focus on political detachment users.
We consider ``political detachment users'' to be users satisfying the following criteria:
they were not following any political party and did not retweet any tweets of political parties.
There were 13,174,064 political detachment users, or 78.7\% of the 16,742,073 users selected for this study.
It is important that political parties deliver information to political detachment users because the amount of them is large.

\section{Results and Discussion}

\subsection{RQ1: Homogeneity of Followers}

Homogeneity is measured by the degree of overlap between followers of a political party and their followers.
This section addresses the following research questions:
who is following the user who is following a political party account?
Political parties can deliver information to their followers, but cannot deliver to users who are not followers.
If the followers retweet the tweets of political parties, the parties can deliver information to users other than their followers.
Therefore, who the follower of a follower is will influence information diffusion power.
If followers of users following a political party are composed only of followers of the political party,
even if the followers retweet the tweets of the political party, the party delivers the information only to the followers.

\begin{figure}[tp]
  \centering
  \includegraphics[width=0.83\linewidth]{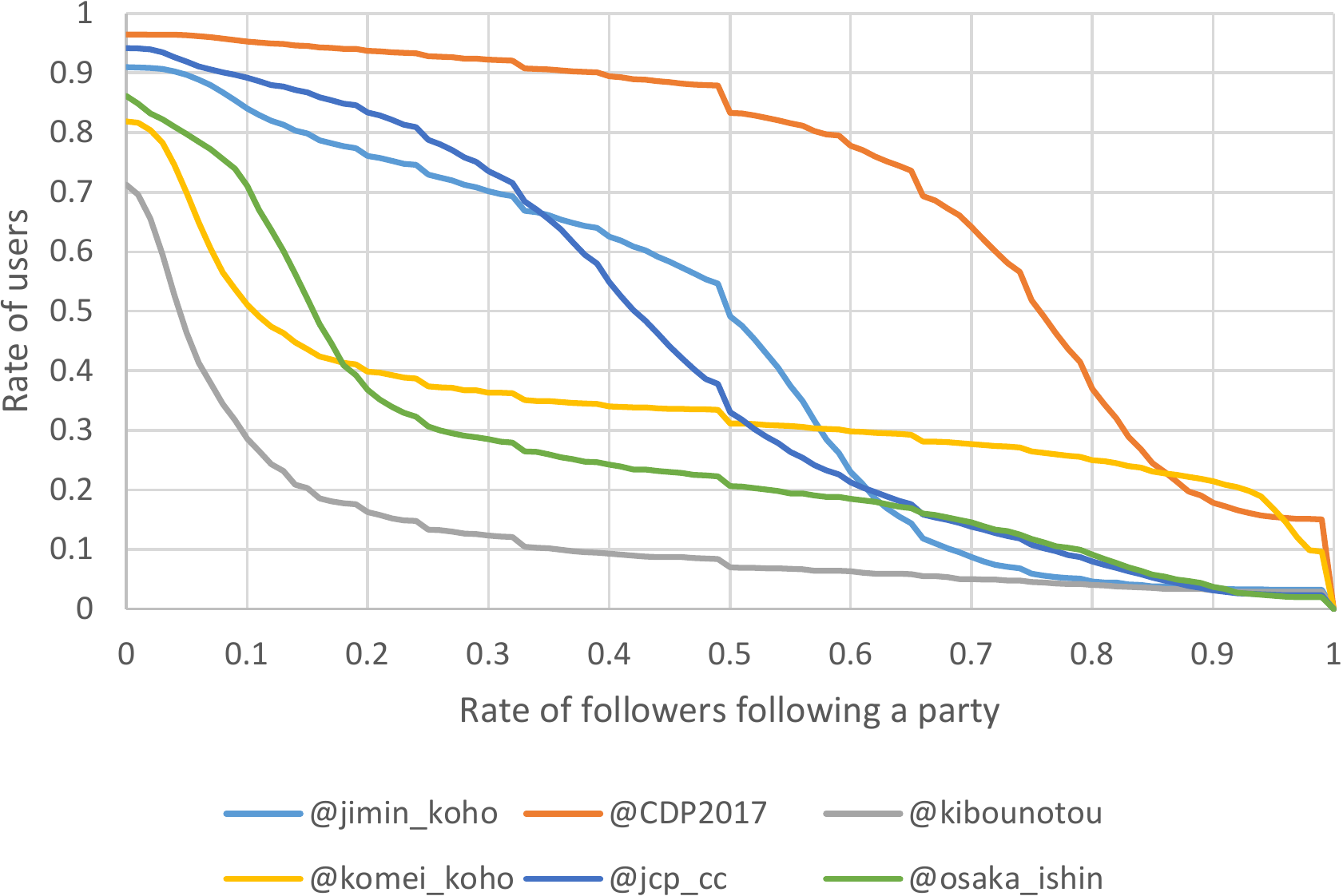}
  \caption{Homogeneity of followers: the x-axis indicates the rate of followers following a political party. For example, if a user has ten followers, of which six followers follow a political party, then the x-axis points to 0.6. The y-axis indicates the rate of such users in the followers of a political party. Homogeneity is higher at the upper right, and homogeneity is lower at the lower left. In the followers of users following @CDP2017, 50\% of the followers are users whose followers follow @CDP2017 by over 75\%. In @kibounotou, 50\% of the followers are users whose followers follow @kibounotou by over 4\%.}
  \label{fig:RQ1}
\end{figure}

Figure~\ref{fig:RQ1} shows the homogeneity of the followers of each of the political parties.
Homogeneity is higher at the upper right, and homogeneity is lower at the lower left.
In the followers of users following @CDP2017, 50\% of the followers are users whose followers follow @CDP2017 by over 75\%.
If homogeneity is high, we can consider that the community of supporters (followers) was made sufficiently because the subgraph becomes dense.
A new party, @CDP2017, succeeded in building a community, but the second new party @kibounotou failed.

\begin{figure}[tp]
  \centering
  \includegraphics[width=0.83\linewidth]{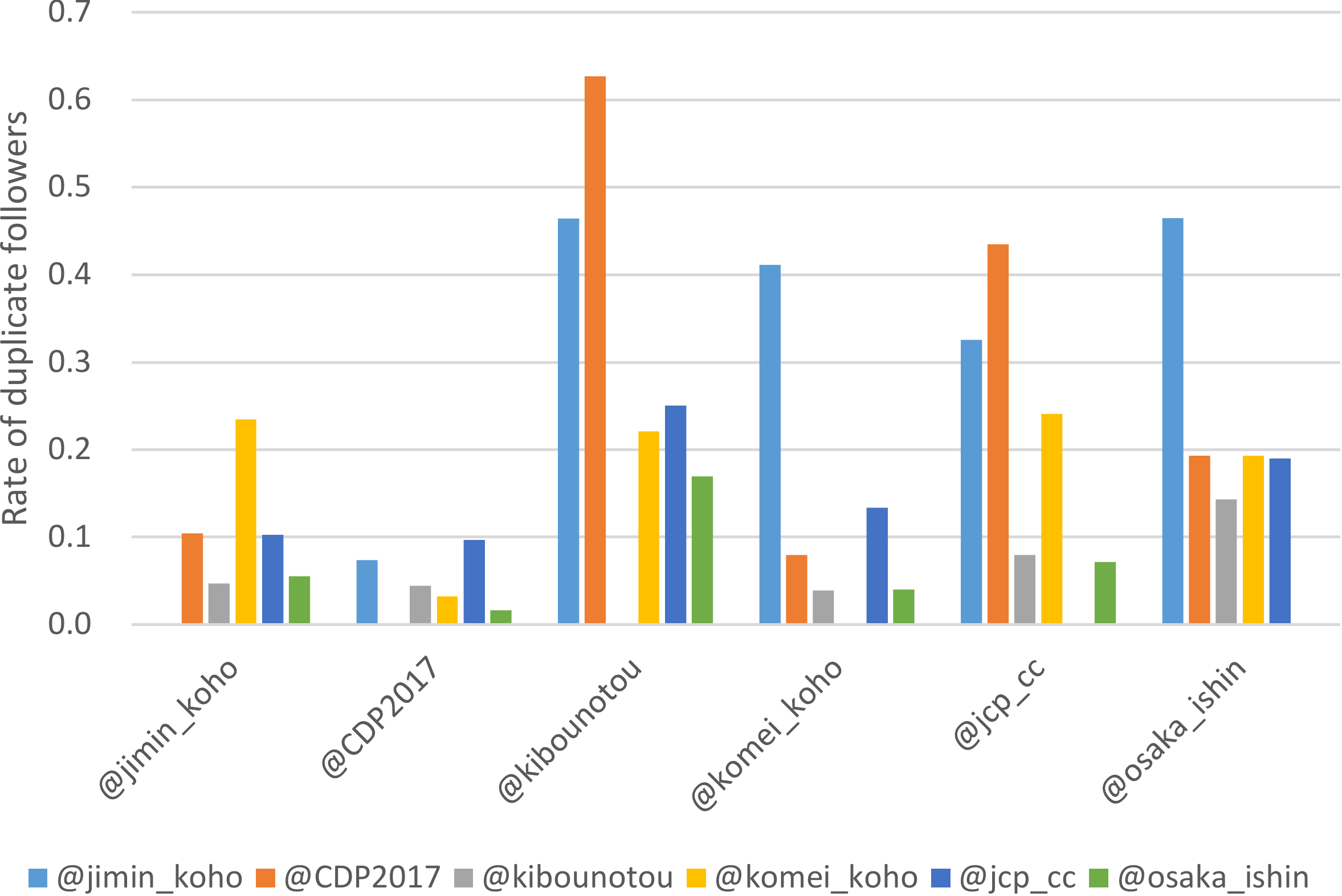}
  \caption{Duplicate followers: the y-axis indicates the rate of followers of the x-axis following a political party. The followers of @CDP2017 typically do not follow other political parties. 63\% of the followers of @kibounotou are following @CDP2017. Followers with low homogeneity tend to follow other political parties.}
  \label{fig:RQ1-2}
\end{figure}

A highly homogeneous group of followers tends to not follow other political parties.
Figure~\ref{fig:RQ1-2} shows the rate of the duplicate followers for each political party.
The followers of @CDP2017 typically do not follow other political parties.
Such users will only receive information from a specific party.
Therefore, the political parties can deliver their information to such users intensively.

\subsection{RQ2: Activity Level of Followers} \label{sec:RQ2}

The activity level is mainly measured by the rate of followers who retweeted.
This section addresses the following research questions:
how many tweets were retweeted by followers of a political party account?
For political parties to deliver information to many users, the followers need to be frequently retweeted because many users are political detachment users.
If any followers of a political party did not retweet, the political party can deliver information only to the followers.
Even if tweets are delivered to users by retweets, users do not necessarily see it~\cite{Comarela2012,Rodriguez2014}.
In this analysis, we focus on whether political parties ``can deliver'' information, so we do not consider whether users actually saw the tweets.

\begin{figure}[tp]
  \centering
  \includegraphics[width=0.9\linewidth]{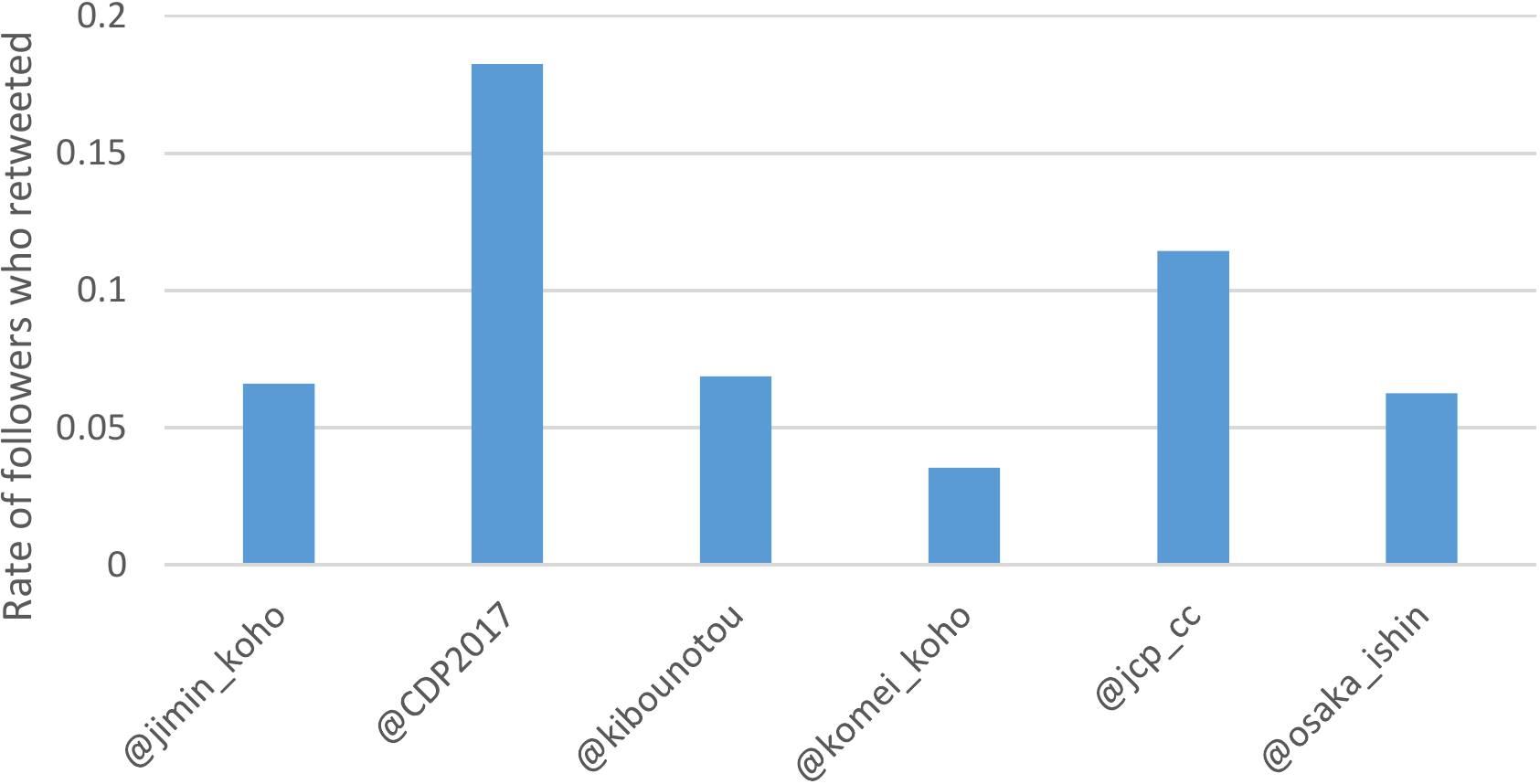}
  \caption{Activity level of followers: how many followers retweeted. 18\% of the followers of @CDP2017 retweeted the tweets of @CDP2017. The tweets of @kibounotou were retweeted by the non-followers.}
  \label{fig:RQ2}
\end{figure}

Figure~\ref{fig:RQ2} shows the activity level of the followers of each of the political parties.
In @CDP2017, 18\% of the followers retweeted the tweets of @CDP2017.
This value is higher than the values of other political parties.
@CDP2017 was created on 2 October 2017,
and there are no followers who have not used Twitter for a long time.
Thus, if the activity level of followers was high, the followers read the tweets and may have frequently retweeted.
However, the tweets of @kibounotou were retweeted by non-followers.
The homogeneity of the followers of @kibounotou is low, and @kibounotou can deliver information to various users, and as a result, non-followers may have frequently retweeted.

\subsection{RQ3: Information Diffusion Power}

Information diffusion power is mainly measured by the number of users who were delivered the tweets.
This section addresses the following research questions:
how many tweets were delivered to political detachment users?
The number of times a user was delivered the tweets of a political party is calculated as follows:
if the user is a follower of the political party, number of the tweets of political parties,
else number of times that followees (followed users) of the user retweeted the tweets of the political party.
We assume that a user was ``delivered'' the tweets of a political party because the tweet appears on the timeline of the user.

\begin{figure}[tp]
  \centering
  \includegraphics[width=0.9\linewidth]{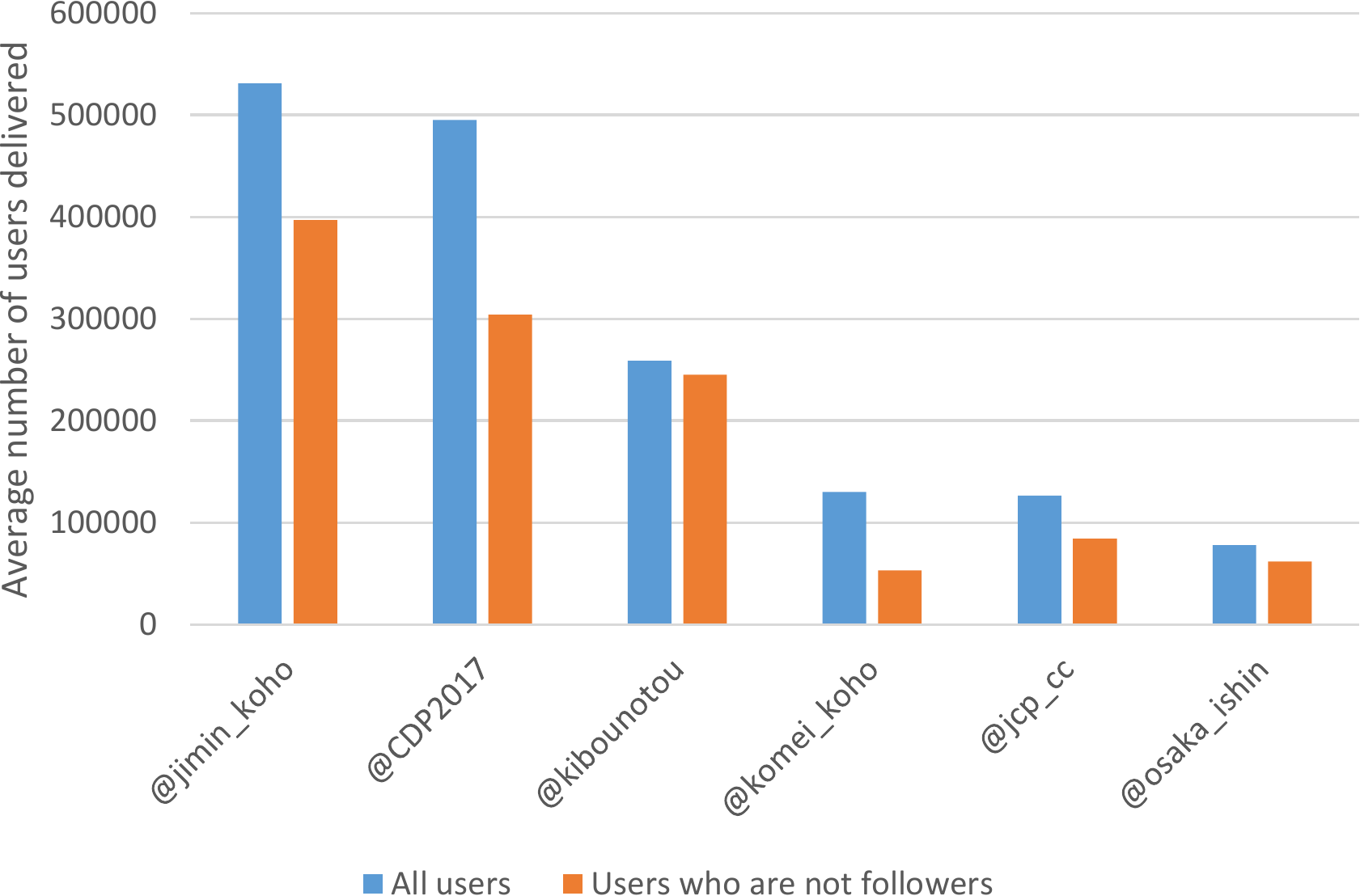}
  \caption{Information diffusion power: the power of @jimin\_koho was the strongest. @CDP2017 was the second strongest, despite having the largest number of followers. @komei\_koho was weak in its power to deliver tweets to non-followers.}
  \label{fig:RQ3}
\end{figure}

Figure~\ref{fig:RQ3} shows the information diffusion power of each of the political parties.
The picture of the information diffusion power is the average number of users who are delivered a tweet by a political party.
If the homogeneity is low and activity level is high, the information diffusion power will also be higher.
The power of @CDP2017 was the second best, despite having the largest number of followers.
The reason is that the homogeneity of the followers is high.
Specifically, there is no diversity for followers of the followers of the political party.
Clearly, the information diffusion power cannot be measured only by the number of followers.
The power of @jimin\_koho, which is a ruling party, is high.
However, @komei\_koho (Komeito) was weak in its power to deliver tweets to non-followers, despite being a ruling party.

\section{Conclusion}

We performed a detailed analysis of social graphs and users who retweeted tweets of political parties during the 48th general election for Japan's Lower House in 2017.
Our aim is to obtain accurate information regarding the diffusion power for each party rather than just the number of followers.
To this end, we addressed three research questions.
The results indicate that a user following a user who follows a political party account tended to also follow the account.
This means that it does not increase diversity because users who follow each other tend to share similar values.
We also found that followers of a specific party frequently retweeted the tweets.
However, since users following a user who follows a political party account are not diverse, political parties delivered the information only to a few political detachment users.

\bibliographystyle{splncs04}
\bibliography{socinfo}

\section*{Notices}

This is the author's version of a work that was accepted for publication in the 10th International Conference on Social Informatics (SocInfo 2018).\\
The original publication is available at www.springerlink.com.\\
\url{https://doi.org/10.1007/978-3-030-01159-8\_32}

\section*{Supplement}

\begin{figure}[h]
  \centering
  \includegraphics[width=0.9\linewidth]{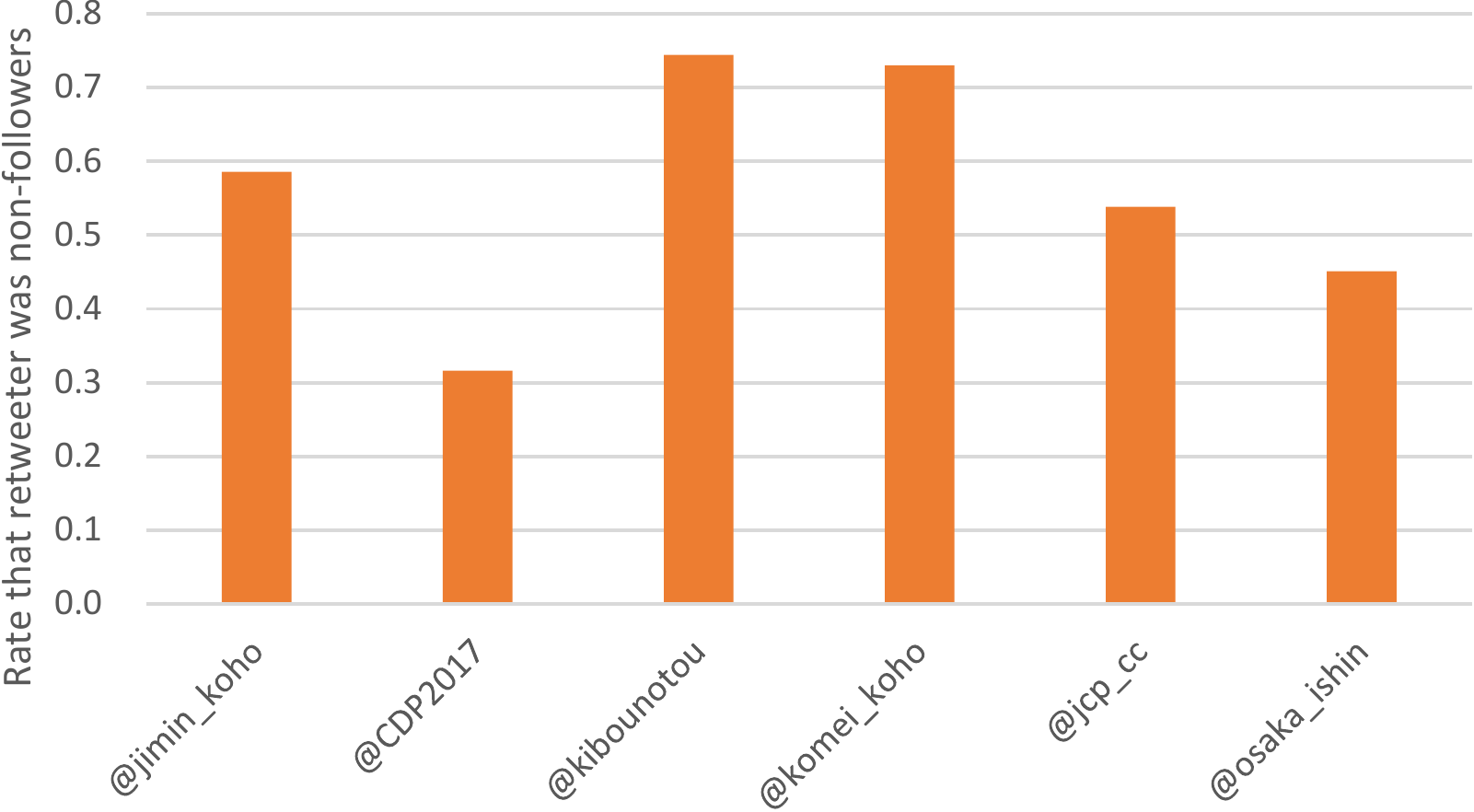}
  \caption{Who was the retweeter of each political party? The tweets of @CDP2017 were retweeted by the followers, but the tweets of @kibounotou were retweeted by the non-followers.}
  \label{fig:RQ2-supplement}
\end{figure}

We add Figure~\ref{fig:RQ2-supplement} to reinforce the second paragraph of Section~\ref{sec:RQ2}.
We forgot to put this figure in the publisher's version.
A political party's retweeter is either a follower or a non-follower of the party.
Figure~\ref{fig:RQ2-supplement} shows the rate of the non-followers in the retweeters for each political party.

\includepdf[pages=-]{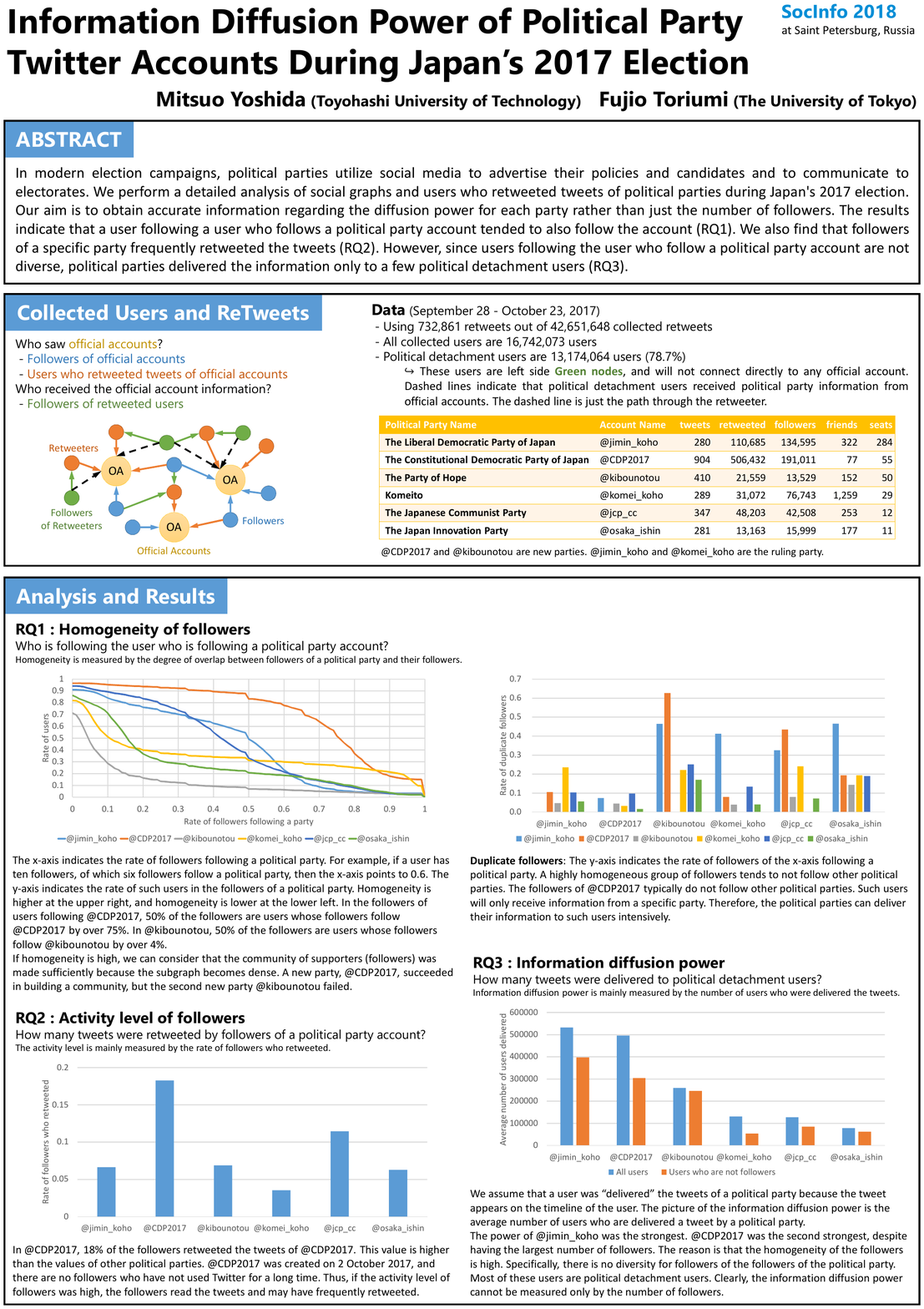}

\end{document}